\documentclass[prl,showpacs,preprintnumbers,keywords,twocolumn]{revtex4-1}
\usepackage{epsfig}
\usepackage{amssymb,amsmath,amsthm}
\usepackage[normalem]{ulem}	
\usepackage{color}
\usepackage[utf8]{inputenc}
\usepackage[colorlinks=true,linkcolor=blue,citecolor=blue]{hyperref}
\usepackage{soul}
\usepackage{braket}
\usepackage[none]{hyphenat}
\usepackage{enumitem}
\usepackage{mathtools,amssymb}

\begin{document}
\sloppy

\title{Lie algebraic solution of the Kratzer oscillator in diatomic molecules}

\author{Boris Maul\'en$^{1}$}
\author{Jos\'e Mauricio Gonz\'alez$^{2}$} 
\vspace{2cm}
\affiliation{$^{1}$ Departamento de Ciencias Qu\'imicas\\ 
$^{2}$ Departamento de Ciencias F\'isicas, \\
Facultad de Ciencias Exactas, Universidad Andr\'es Bello, Santiago, Chile.}

\begin{abstract}
The study of diatomic molecules plays a central role in the understanding of the chemical bond. For their simplicity, they serve as a model for the study of more complex molecular systems. In this article, we solve the rovibrational Schr\"odinger equation for diatomic molecules using the Kratzer oscillator, by means of \textbf{so(2,1)} Lie algebra. The energies and bound states for this simple model are obtained through a canonical transformation of the molecular Hamiltonian. The main contribution of the Lie-algebraic approach is that this allows us to reduce the degree of Schr\"odinger equation, obtaining a first-order differential equation whose resolution is considerably simpler than the original one. Additionally, we give the physical insight of the symmetry transformation of the SO(2,1) Lie group and show the relationship between this group and its associated Lie algebra. Finally, as an illustrative example, we calculated the selection rules for the vibrational quantum number by the use of transformation rules of SO(2,1) Lie group.
\end{abstract}


\maketitle

\section{I. Introduction}
Diatomic molecules are the simplest physical systems that present a covalent chemical bond. Comprehension of these small molecules serves as building blocks for the chemical bonding theory. For example, at the beginning of quantum chemistry, the hydrogen-ion molecule $H_2^+$ served as the first application of Schr\"odinger equation to a molecular problem. From the solutions to this problem, it was possible to construct the wave functions for more complex molecules taking appropriate linear combinations. Also, the degrees of freedom of a molecule can be decomposed into different classes depending on the energies involved, namely, translational, rotational, vibrational, and electronic ones. In particular, the rotation-vibration (rovibration) motion of molecules is associated with rotational and vibrational spectroscopy. It is for this reason that the study of diatomics is not only relevant for theoretical chemistry but also experimental molecular physics and astrochemistry \cite{astrochemistry}, \cite{database}.

A realistic description for molecular rovibration must be carried out using rigorous quantum mechanics. The stationary Schr\"odinger equation for this problem is given by
\begin{equation}
\left[\dfrac{p^2_{r}}{2\mu}+\dfrac{l(l+1)\hbar^{2}}{2\mu r^2}+U(r) \right]\ket{R_{v,l}} =E_{v,l}\ket{R_{v,l}}\label{sch_eq},
\end{equation}
where $E_{v,l}$ and $\left\lbrace \ket{R_{v,l}}\right\rbrace $ are the energies and the rovibration states of the diatomic molecule, which are labeled by both vibrational quantum number $v$ and the rotational quantum number $l$, $r$ is the internuclear distance, $p_r$ is the radial component of the momentum and $\mu$ is the reduced mass of the molecule. Depending on the nature of potential function $U(r)$, the Hamiltonian spectrum will be purely discrete or will be formed by the union of its discrete and continuum parts, and, only in few cases, Eq. $\left( \ref{sch_eq}\right)$ has an analytical solution.

If our attention is focused on the rovibrational behavior of the molecule close to the internuclear equilibrium distance $r_e$, the harmonic oscillator and rigid rotor approximations works. However, for excited rovibrational states far from ground state, these approximations are not applicable because they are not considering effects due to \textit{anharmonicity}. On the other hand, the Born-Oppenheimer (BO) curve for diatomic molecules exhibit a typical asymmetry which is a consequence of the anharmonicity of molecular oscillations in excited vibrational states.

A way to correct the energies and the states given by the harmonic oscillator model is by the use of empirical potential functions. These potential functions depend on experimental parameters, such as spectroscopic constants, allowing the study of the BO curve in all its domain. A well-known empirical potential function is the \textit{Morse potential}. This potential function has the typical profile of the experimental BO curve. Additionally, Eq. $\left(\ref{sch_eq} \right)$ with this potential can be solved analytically. However, Morse potential function is a merely arbitrary function, and it does not have theoretical foundations. Fortunately, through the use of molecular virial theorem \cite{rioux}, \cite{virial2} it is possible to find a differential equation whose solution correspond to a general internuclear potential function $U\left( r\right)$,
\begin{equation}
r^{2}\dfrac{d^{2}U(r)}{dr^{2}}+2r\dfrac{dU(r)}{dr}+2U(r)=W(r)\label{EDO_U},
\end{equation}
where $W(r)$ function is given by
\begin{equation}
W(r)=-\dfrac{1}{r}\dfrac{d}{dr}\left(r^{2}\left\langle T_{el}\right\rangle  \right),
\end{equation}
where $\left\langle T_{el}\right\rangle$ is the average of electronic kinetic energy. Eq. (\ref{EDO_U}) was deduced for the first time by Borkman and Parr  \cite{parr}, and it has a general solution given by
\begin{eqnarray}
\nonumber U(r)=\dfrac{1}{r}\left[2r_{e}U_{e}+\int_{r_{e}}^{r}{W(r')dr'} \right]+\\
\dfrac{1}{r^{2}}\left[r^{2}_{e}U_{e}-\int_{r_{e}}^{r}{r'W(r')dr'} \right]\label{sol_U}
\end{eqnarray}
where $U_{e}\equiv U(r_{e})$ correspond to the value of the potential function at the internuclear distance $r_e$ and it is the negative of equilibrium dissociation energy $D_e$. The solution (\ref{sol_U}) shows that an internuclear interaction potential that satisfies the virial theorem can have several forms according to the specific form of the inhomogeneity $W(r)$. The simplest solution is 
\begin{equation}
U(r)=U_{e}\left[2\left( \dfrac{r_{e}}{r}\right) -\left( \dfrac{r_{e}}{r}\right)^{2}\right], \label{kratzer}
\end{equation}
which is obtained when $W(r)=0$ for all internuclear distance. Under this condition, we obtain for the electronic kinetic energy
\begin{equation}
\left\langle T_{el}\right\rangle=\left(\dfrac{r_{e}}{r} \right)^{2}\label{particle_box}.
\end{equation}
The potential function (\ref{kratzer}) is the well-known \textbf{Kratzer oscillator} \cite{kratzer}. Kratzer oscillator is an anharmonic internuclear potential that presents a correct asymptotic behavior when $r\rightarrow 0 $ and when $r\rightarrow\infty$, and it has discrete spectrum for all $E_{n,l}<D_{e}$, and continuum one for $E_{n,l}>D_{e}$. Moreover, Kratzer oscillator has shown to be a successful model that allows the reproduction of the spectroscopic constants of diatomic molecules \cite{hooydonk}-\cite{kratzer_H2}. Note that the equation (\ref{particle_box}) has the form of the energy of a particle confined in a square well, i.e., it is proportional to the square reciprocal of the well length. Thus, Kratzer oscillator must be fulfilled when $\left\langle T_{el}\right\rangle $ behaves like the energy of the particle in a square well. Kratzer oscillator have central importance in theoretical chemistry due to the possibility to study bound-continuum transitions, which are responsible for the rupture of chemical bonds.

There are some works concerning about solving Eq. (\ref{sch_eq}) through traditional methods of differential equation \cite{matamala}, or by means of factorization method of ladder operators \cite{oyewumi}-\cite{ladder_SU2}. Despite the fact that these methods gives the exact solution to the problem, in the case of the hypergeometric series method provided by A. R. Matamala, the mathematical complexity becomes overwhelming, while in the case of factorization method, there is no clear insight of the physical meaning of the ladder operators nor it shows the relationship between the Lie algebra structure with the corresponding Lie group. In this work, we show an alternative solution method based on the Lie algebra $\mathbf{so(2,1)}$ whose main advantages of the above methods is its simplicity which we obtain the energies and the bound states, and that allows us to reduce the complexity of Eq. $\left(\ref{sch_eq}\right)$ for the Kratzer oscillator, giving a first-order differential equation. Additionally, we give the physical insight of the generators of this algebra, relating with the symmetry transformations of the $SO(2,1)$ Lie group.

\section{II. Radial representation of $\mathbf{so(2,1)}$ Lie algebra}
To solve Eq. (\ref{sch_eq}) for Kratzer oscillator, it is necessary to work with a Lie algebra that reproduces the physics of the problem, i.e., with an eigenvalue spectrum with only a lower bound (unlike of $\mathbf{so(3)}$ Lie algebra, which has both lower and upper bounds). Thus, we can relate the lower bound of the algebra with the ground state of the Kratzer oscillator. $\mathbf{so(2,1)}$ Lie algebra fulfils these requirements. Generators of $\mathbf{so(2,1)}$ are $T_1, T_2$, and $T_3$, which satisfy the commutation relations
\begin{equation}
\left[ T_{1},T_{2}\right] =-i\hbar T_{3}, \hspace{0.4cm}
\left[ T_{2},T_{3}\right]=i\hbar T_{1}, \hspace{0.4cm}
\left[T_{3},T_{1}\right]=i\hbar T_{2}.\label{commutation_rules}
\end{equation}
Then, the molecular Hamiltonian is a function of the set of observables $r$ and $p_{r}$. Thus, it is necessary to write the generators of $\mathbf{so(2,1)}$ in terms of the observables position and radial momentum. An appropriate representation is given by \cite{cooke}
\begin{equation}
T_{1}=\dfrac{1}{2}\left[ \dfrac{1}{a^2} R^{2-a} P^2+\tau R^{-a}-R^{a}\right],\label{T1}
\end{equation}
\begin{equation}
T_{2}=\dfrac{1}{a}\left[R P-\dfrac{i}{2}(a-1)\hbar \right],\label{T2}
\end{equation}
\begin{equation}
T_{3}=\dfrac{1}{2}\left[ \dfrac{1}{a^2} R^{2-a} P^2+\tau R^{-a}+R^{a}\right],\label{T3}
\end{equation}
where $R$ and $P$ are generalized observables which must fulfill the canonical commutation relation, $\left[ R,P\right]=i\hbar$. In Eqs. (\ref{T1})-(\ref{T3}) the parameter $a$ fits according to the potential $U(r)$, and $\tau$ contains the specific information of the system such as spectroscopic constants and quantum numbers. From generators of $\mathbf{so(2,1)}$, it is possible to define the \textit{Casimir} operator,
\begin{equation}
T^{2}=-T^{2}_{1}-T^{2}_{2}+T^{2}_{3}.\label{casimir}
\end{equation}
Then, we choose $T^{2}$ and $T_{3}$ to represent the physics of the system because these are Hermitian and commute to each other. Thus, there are simultaneous eigenstates $\ket{Q,q_{n,l}}$ for both operators, and the respective eigenvalues are real numbers,
\begin{equation}
T^{2}\ket{Q,q_{n,l}}=Q\ket{Q,q_{n,l}},
\end{equation}
\begin{equation}
T_{3}\ket{Q,q_{n,l}}=q_{n,l}\ket{Q,q_{n,l}}, \label{eigen_T3}
\end{equation}
where $Q$ and $q_{n,l}$ are the eigenvalues of $T^{2}$ and $T_{3}$, whose labels $n$ and $l$ will get related to the quantum numbers of the rovibrational states of the molecule.

\section{III. Ladder operators of $\mathbf{so(2,1)}$}
It is possible to define ladder operators for $\mathbf{so(2,1)}$ Lie algebra from their generators as
\begin{equation}
T_{\pm}=T_{1}\pm i T_{2}.
\end{equation}
Some important relationships are \cite{cooke}
\begin{equation}
\left[ T_{3}, T_{\pm}\right] = \pm \hbar T_{\pm},\hspace{0.35cm}
\left[ T_{+}, T_{-}\right]=-2 \hbar T_{3}. \label{ladder_commutator}  
\end{equation}
Then, it is possible to express the Casimir operator of $\mathbf{so(2,1)}$ in terms of these operators as
\begin{equation}
T^{2}=-T_{\pm} T_{\mp}+\left( T_{3}\mp \hbar \right) T_{3}.\label{casimir2} 
\end{equation}
The meaning of ladder operators is obtained by the effect on the eigenvalues equation of $T_{3}$ and $T^{2}$. Thus, the action of $T_{\pm}$ on the eigenvalues equation of $T_{3}$ is given by
\begin{equation}
T_{\pm} T_{3} \ket{Q,q_{n,l}} = q_{n,l} T_{\pm} \ket{Q,q_{n,l}}.
\end{equation}
It is very important to note that ladder operators of $\mathbf{so(2,1)}$ affects only the label $n$ without affecting $l$. Thus, using the first equation of (\ref{ladder_commutator}), we found that
\begin{equation}
T_{3} T_{\pm} \ket{Q,q_{n,l}} = \left(q_{n,l} \pm \hbar\right) T_{\pm} \ket{Q,q_{n,l}}.
\end{equation}
Lower bound of $T_{3}$ is defined by $T_{-}\ket{Q,q_{0,l}}=0$, and the respective eigenvalue equation is
\begin{equation}
T_{3} \ket{Q,q_{0,l}}=q_{0,l} \ket{Q,q_{0,l}},
\end{equation}
where $q_{0,l}$ is the eigenvalue for the lower bound, and whose relationship with $q_{n,l}$ is given by means of the $nth$-power of raising operator, i.e.,
\begin{equation}
T_{3} T^{n}_{+} \ket{Q,q_{0,l}} = \left(q_{0,l}+n\hbar \right) \ket{Q,q_{0,l}},\label{T3T+n}
\end{equation} 
with $q_{n,l}=q_{0,l}+n\hbar$. In this way, it is possible to obtain the eigenvalue for the $nth$-excited state from $q_{0,l}$.

Useful relationship between $Q$ and $q_{0,l}$ arise from the action of $T_{+}T_{-}$ on $\ket{Q,q_{0,l}}$. Using  Eq. (\ref{casimir2}) we have,
\begin{eqnarray}
\nonumber T_{+}T_{-}\ket{Q,q_{0,l}}=\left(-T^{2}+T^{2}_{3}-\hbar T_{3} \right)\ket{Q,q_{0,l}}\\
=\left( -Q+q^{2}_{0,l}-\hbar q_{0,l}\right)\ket{Q,q_{0,l}}=0.  
\end{eqnarray}
It follows that
\begin{equation}
Q=q_{0,l}\left(q_{0,l}-\hbar \right).\label{Q_q}
\end{equation}

\section{IV. Energies of Kratzer oscillator}
Eq. (\ref{sch_eq}) for Kratzer oscillator is given by
\begin{equation}
\left[ \dfrac{p^{2}_{r}}{2\mu}+\dfrac{l(l+1)\hbar^{2}}{2\mu r^{2}}+\dfrac{\alpha}{r}+\dfrac{\beta}{r^2}\right]\ket{R_{v,l}}=E_{v,l} \ket{R_{v,l}}, \label{sch_kratzer}
\end{equation}
with $\alpha=-2D_{e} r_{e}$ and $\beta=D_{e} r^{2}_{e}$. It is possible to scale Eq. (\ref{sch_kratzer}) obtaining thus
\begin{eqnarray}
\nonumber \dfrac{1}{2}\left\lbrace  \sigma r p^{2}_{r}+\left[ l(l+1)\hbar^{2}+2\mu \beta\right]\dfrac{\sigma}{r}-2\mu\sigma r E_{v,l} \right\rbrace  \\
\ket{R_{v,l}}=-\mu \sigma \alpha \ket{R_{v,l}},\label{sch_alg}
\end{eqnarray}
where $\sigma$ is a scaling factor that allows us to perform a canonical transformation over the observables  $r$ and $p_r$. In this way, it will be possible to express the molecular Hamiltonian  $H$, which is a function of $r$ and $p_r$, in terms of some generators of $\mathbf{so(2,1)}$. On the other hand, generators of this Lie algebra are function of another set of generalized observables, $R$ and $P$, which fulfils $\left[ R , P \right]=i\hbar $. So, the relationship between the sets of observables $\left\lbrace r,p_{r}\right\rbrace $ and $\left\lbrace R,P\right\rbrace $, and, therefore, the transformation of $H$ by the scaling factor $\sigma$, must be preserve the canonical commutation relation. The above is achieved only if $R=\dfrac{r}{\sigma}$ and $P=\sigma p_{r}$. In this transformation, when changing the observables $\left\lbrace r,p_{r}\right\rbrace $ for $\left\lbrace R,P\right\rbrace $ in $H$, will be affected the states $\left\lbrace \ket{R_{v,l}}\right\rbrace $ and the energies $E_{v,l}$. Considering this, Eq. (\ref{sch_alg}) takes the form
\begin{small}
\begin{eqnarray}
\nonumber \dfrac{1}{2}\left\lbrace R P^{2}+\left[ l(l+1)\hbar^{2}+2 \mu \beta \right]R^{-1}-2\mu \sigma^{2} \varepsilon_{n,l} R \right\rbrace \ket{\bar{R}_{n,l}} \\
=-\mu \sigma \alpha \ket{\bar{R}_{n,l}},\label{sch_alg2}
\end{eqnarray}
\end{small}
\hspace{-0.36cm} where $\left\lbrace \ket{\bar{R}_{n,l}}\right\rbrace $ are the transformed states of $\left\lbrace \ket{R_{v,l}}\right\rbrace $ associated with the algebraic Hamiltonian $H_{A}$ defined as
\begin{small}
\begin{equation}
H_{A}=\dfrac{1}{2}\left\lbrace R P^{2}+\left[ l(l+1)\hbar^{2}+2 \mu \beta \right] R^{-1}-2 \mu \sigma^{2} \varepsilon_{n,l} R \right\rbrace,\label{alg_ham}
\end{equation}

\end{small}
\hspace{-0.39cm} and $\varepsilon_{n,l}$ are the scaled energies. Comparing $T_{3}$ of $\mathbf{so(2,1)}$ for $a=1$,
\begin{equation}
T_{3}=\dfrac{1}{2}\left(R P^{2}+\tau R^{-1}+R \right), 
\end{equation}
with Eq. (\ref{alg_ham}), we obtain
\begin{equation}
\tau=l(l+1)\hbar^{2}+2 \mu \beta,
\end{equation}
\begin{equation}
\varepsilon_{n,l}=-\dfrac{1}{2 \mu \sigma^{2}}.\label{energies}
\end{equation}
On the other hand, eigenvalue equation for $H_{A}$ is
\begin{equation}
H_{A}\ket{\bar{R}_{n,l}}=-\mu \sigma \alpha \ket{\bar{R}_{n,l}},
\end{equation}
and by comparison with Eq. (\ref{eigen_T3}), we obtain
\begin{equation}
q_{n,l}=-\mu \sigma \alpha.\label{qn}
\end{equation}
Eq. (\ref{energies}) tells us that the scaled energies are determined by the scaling factor $\sigma$, while Eq. (\ref{qn}) restricts the possible values of $\sigma$ to the values of $q_{n,l}$. Also, $q_{n,l}$ is obtained from the lower bound eigenvalue $q_{0,l}$ through $q_{n,l}=q_{0,l}+n\hbar$. Analytical expression for $q_{0,l}$ exists, and it is given by \cite{cooke}
\begin{equation}
q_{0,l}=\dfrac{\hbar}{2}\left( 1\pm \sqrt{\dfrac{4 \tau}{\hbar^{2}}+1}\right).\label{q0}
\end{equation}
Note that $q_{n,l}$ and also $q_{0,l}$ depends on the rotational quantum number $l$ through $\tau$. Then, replacing $\tau$ in Eq. (\ref{q0}), and taking the positive root, we found $q_{0,l}$ for Kratzer oscillator:
\begin{equation}
q_{0,l}=\hbar \left[ \dfrac{1}{2}+ \sqrt{\left(l+\dfrac{1}{2} \right)^{2}+\dfrac{2 \mu \beta}{\hbar^{2}}}\right]. \label{q0_kratzer}
\end{equation}
Then, using
\begin{equation}
q_{0,l}+n\hbar=-\mu \sigma \alpha,
\end{equation}
we obtain
\begin{equation}
\sigma =-\dfrac{\hbar}{\mu \alpha} \left[  n+ \dfrac{1}{2}+\sqrt{\left( l+\dfrac{1}{2}\right)^{2}+\dfrac{2 \mu \beta}{\hbar^{2}}} \right] . \label{sigma} 
\end{equation}
By last, the energy eigenvalues are obtained replacing $\sigma^2$  in Eq. (\ref{energies}): 
\begin{equation}
\varepsilon_{n,l}=-\dfrac{2 D^{2}_{e} r^{2}_{e} \mu}{\hbar^{2}\left[  n+\dfrac{1}{2}+\sqrt{ \left(l+\dfrac{1}{2} \right)^{2}+\dfrac{2 D_{e} r^{2}_{e} \mu}{\hbar^{2}} } \right] ^{2} }.
\end{equation}
Therefore, the rovibration energies for Kratzer oscillator depends only on the spectroscopic constant $D_e$, $r_e$ as well as on the reduce mass of the diatomic molecule and on the vibrational and rotational quantum numbers $n$ and $l$. We take the positive root in $q_{0,l}$ because, as long as the molecule takes higher rovibrational states (high values of $n$ and $l$), the energies must be increased. This is not achieved with the negative root.

\section{V. Bound states of Kratzer oscillator}
In order to obtain the bound states of Kratzer oscillator, we consider the effect of ladder operators $T_{\pm}$ on the basis  $\left\lbrace \ket{Q,q_{n,l}}\right\rbrace $, i.e., 
\begin{equation}
T_{\pm} \ket{Q,q_{n,l}}=c_{\pm} \ket{Q,q_{n\pm 1,l}},
\end{equation}
with $q_{n\pm 1,l}\equiv q_{n,l}\pm \hbar$. Then, we need to evaluate the constant $c_{\pm}$. For this, it is necessary to calculate the average of $T_{\pm}T_{\mp}$ by means of Eq. (\ref{casimir2})
\begin{eqnarray}
\nonumber \left\langle T_{\pm} T_{\mp}\right\rangle = \bra{Q,q_{n,l}} \left(-T^{2} +T^{2}_{3}\pm \hbar T_{3} \right) \ket{Q,q_{n,l}}\\
=\left(-Q+q^{2}_{n,l}\pm \hbar q_{n,l} \right), \label{T+-T-+}
\end{eqnarray}
where we have supposed that the basis  $\left\lbrace \ket{Q,q_{n,l}}\right\rbrace $ is normalized.
On the other hand, considering that $T_{\pm}=T^\dagger_{\mp}$, we obtain
\begin{equation}
\left\langle T_{\mp} T_{\pm} \right\rangle =\braket{T_{\pm}Q,q_{n,l}|T_{\pm}Q,q_{n,l}}=\Vert\ket{T_{\pm}Q,q_{n,l}}\Vert^2,\label{mean_value}
\end{equation}
and by comparison with (\ref{T+-T-+}):
\begin{equation}
c_{\pm}=\sqrt{-Q+q_{n,l}\left(q_{n,l}\pm \hbar \right)}.
\end{equation}
Hence, the action of $T_{\pm}$ on the basis $\left\lbrace \ket{Q,q_{n,l}}\right\rbrace$ is written as
\begin{equation}
T_{\pm} \ket{Q,q_{n,l}}=\sqrt{-Q+q_{n,l}\left(q_{n,l}\pm \hbar \right)  } \ket{Q,q_{n\pm 1,l}}.
\end{equation}
With the last, it is possible to find a recurrence relation that connects the lower bound of $T_3$ with its $nth$-excited state. Applying $T_+$ on $\ket{Q,q_{0,l}}$ we have first
\begin{equation}
T_{+} \ket{Q,q_{0,l}}=\sqrt{-Q+q_{0,l}(q_{0,l}+\hbar)}\ket{Q,q_{1,l}}.
\end{equation}
Then, for $T_{+}^2$ we have
\begin{small}

\begin{equation}
\hspace{-6cm}\nonumber T^{2}_{+}\ket{Q,q_{0,l}} =
\end{equation}
\begin{equation}
 \sqrt{-Q+q_{0,l}(q_{0,l}+\hbar)} \sqrt{-Q+(q_{0,l}+\hbar)(q_{0,l}+2\hbar)}\ket{Q,q_{2,l}},
\end{equation}
\end{small}
and so on.  Therefore, for $T_{+}^n$, we obtain
\begin{small}
\begin{equation}
T^{n}_{+}\ket{Q,q_{0,l}}= \prod_{k=0}^{n-1} \sqrt{-Q+(q_{0,l}+k\hbar)(q_{0,l}+(k+1)\hbar)}\ket{Q,q_{n,l}}. 
\end{equation}
\end{small}
\hspace{-0.36cm} Thus, the $nth$-excited state is written as
\begin{small}
\begin{equation}
\ket{Q,q_{n,l}}=\dfrac{ T^{n}_{+}\ket{Q,q_{0,l}}}{\prod_{k=0}^{n-1} \sqrt{-Q+(q_{0,l}+k\hbar)(q_{0,l}+(k+1)\hbar)}}.\label{states}
\end{equation}
\end{small}
Previously, we showed that while the molecular Hamiltonian $H$ is transformed by a scaling factor $\sigma$, its states $\left\lbrace\ket{R_{v,l}} \right\rbrace $ must be replaced by transformed states $\left\lbrace \ket{\bar{R}_{n,l}}\right\rbrace $, that are directly related to the basis $\left\lbrace \ket{Q,q_{n}}\right\rbrace $. Mathematically, the scaling of the variables $r$ and $p_r$ to the new variables $R$ and $P$ allows us to construct a new operator $H_{A}$ which acts only on a region of the space of states of the original system. Then, a problem will be solved algebraically, only if this region of the space coincides with the subspace spanned by the eigenstates of some generator of a particular Lie algebra. In this case, the subspace spanned by the basis kets of $T_3$ $\left\lbrace \ket{Q,q_{n,l}}\right\rbrace $ of $\mathbf{so(2,1)}$ coincides with the region which acts $H_{A}$.

Physically, the states of the basis $\left\lbrace \ket{Q,q_{n,l}}\right\rbrace $ corresponds precisely to the rovibration bound states of the molecule, that is, those associated only of the discrete part of the molecular Hamiltonian. Considering that the ladder operators $T_{\pm}$ of $\mathbf{so(2,1)}$ only allows to moving through different rovibrational states of the molecule varying only the vibrational quantum number $n$, we can interpret the recurrence relation (\ref{states}) as a relation that connects two different vibrational states for the same rotational state where, in this case, $n$ corresponds to the vibrational quantum number $v$ but associated only to the bound states of the oscillator.
\section{VI. Vibrational wave functions of Kratzer oscillator}
\subsection{A. Vibrational ground state wave function}
In order to obtain the vibrational ground state wave function, we consider conveniently the action of $\left(T_{-}-T_{3} \right)$ on the lower bound $\ket{Q,q_{0,l}}$, i.e.,
\begin{equation}
\left(T_{-}-T_{3} \right)\ket{Q,q_{0,l}}=-q_{0,l}\ket{Q,q_{0,l}}.
\end{equation}
Then, using $T_{-}=T_{1}-iT_{2}$, we have
\begin{equation}
\left(T_1-iT_2-T_3 \right) \ket{Q,q_{0,l}}=-q_{0,l}\ket{Q,q_{0,l}}.
\end{equation}
Replacing $T_1$, $T_2$ and $T_3$ given by Eqs. (\ref{T1})-(\ref{T3}) with $a=1$, we obtain
\begin{equation}
\left( R+i R P-q_{0,l}\right)\ket{Q,q_{0,l}}=0.\label{wave0}
\end{equation}
Projecting Eq. (\ref{wave0}) on the position basis $\left\lbrace \ket{\textbf{r}}\right\rbrace$,
\begin{equation}
\bra{\textbf{r}}R\ket{Q,q_{0,l}}+i\bra{\textbf{r}}R P\ket{Q,q_{0,l}}-q_{0,l}\braket{\textbf{r}|Q,q_{0,l}}=0,
\end{equation}
and expressing it in terms of $r$ and $p_{r}$ (using $R=\dfrac{r}{\sigma}$ and $P=\sigma p_{r}$), it gives
\begin{equation}
\dfrac{r}{\sigma} \braket{\textbf{r}|Q,q_{0,l}}+i r \bra{\textbf{r}}p_{r}\ket{Q,q_{0,l}}-q_{0,l}\braket{\textbf{r}|Q,q_{0,l}}=0.\label{wave1}
\end{equation}
Then, defining $\braket{\textbf{r}|Q,q_{0,l}}\equiv Q_{0,l}(r)Y^{m}_{l}(\theta,\phi)$ and using the representation of the radial component of the momentum $p_{r}$, which is given by
\begin{equation}
\bra{\textbf{r}}p_{r}\ket{Q,q_{0,l}}=-\dfrac{i\hbar}{r}\dfrac{d}{dr}\left[r Q_{0,l}(r)Y^{m}_{l}(\theta,\phi) \right],
\end{equation}
we obtain
\begin{small}

\begin{equation}
\dfrac{r}{\sigma} Q_{0,l}(r)Y^{m}_{l} +i r \hbar \dfrac{d}{dr}\left[r Q_{0,l}(r)Y^{m}_{l} \right]-q_{0,l}Q_{0,l}(r)Y^{m}_{l}=0.
\end{equation}
\end{small}
\hspace{-0.211cm} Here $Q_{0,l}(r)$ is the vibrational ground state wave function. Defining also $f_{0,l}(r)\equiv r Q_{0,l}(r)$ and dividing by $Y^{m}_{l}$, we recast the above equation in the form
\begin{equation}
\dfrac{df_{0,l}(r)}{dr}+\dfrac{1}{\hbar}\left(\dfrac{1}{\sigma}-\dfrac{q_{0,l}}{ r} \right) f_{0,l}(r)=0. \label{gs_wfn_eq40}
\end{equation}
Note that Eq. (\ref{gs_wfn_eq40}) is a first-order differential equation, while the rovibrational Schr\"odinger equation is a second-order one. This reduction on the degree of the Schr\"odinger equation is one of the main advantages of the use of Lie algebras in molecular physics. Thus, the solution of (\ref{gs_wfn_eq40}) is obtained simply by integration:
\begin{equation}
f_{0,l}(r)=Ar^{q_{0,l}/\hbar}e^{-r/\hbar \sigma},
\end{equation}
and hence $Q_{0,l}$ is
\begin{equation}
Q_{0,l}(r)=Ar^{q_{0,l}/\hbar-1}e^{-r/\hbar \sigma}.\label{Q0}
\end{equation}
In Eq. (\ref{Q0}) $q_{0,l}$ is given by (\ref{q0_kratzer}), $\sigma$  (for $n=0$) is defined as
\begin{equation}
\sigma=-\dfrac{q_{0,l}}{\mu \alpha},
\end{equation}
and $A$ is obtained by normalization:
\begin{equation}
A=\left[\int_{0}^{\infty} r^{2q_{0,l}/\hbar}e^{-2r/\hbar\sigma}dr\right]^{-1/2}.
\end{equation}
Also the integral in $A$ is expressed in terms of the Gamma function,
\begin{small}

\begin{equation}
\int_{0}^{\infty} r^{2q_{0,l}/\hbar}e^{-2r/\hbar\sigma}dr = \dfrac{2 q_{0,l}}{\hbar}\left( \dfrac{\hbar\sigma}{2}\right)^{2q_{0,l}/\hbar+1}\Gamma\left(\dfrac{2q_{0,l}}{\hbar} \right).
\end{equation}
\end{small}
\hspace{-0.17cm}Therefore, the vibrational ground state wave function for Kratzer oscillator is given by (in atomic units, $\hbar=1$)
\begin{small}

\begin{equation} 
Q_{0,l}(r)=\dfrac{r^{q_{0,l}-1}e^{\mu \alpha r/ q_{0,l} }}{\sqrt{2 q_{0,l}\left( -\dfrac{ q_{0,l}}{2 \mu \alpha}\right)^{2q_{0,l}+1}\Gamma\left(2q_{0,l} \right)}} \label{Q0_normalizada}
\end{equation}

\end{small}
Depending on the particular value of $l$ it is possible having different expressions for $q_{0,l}$ and hence, several vibrational ground state wave functions for distinct rotational states of the molecule. Thus, the rovibrational ground state ($n=l=0$) is given by Eq. (\ref{Q0_normalizada}) with $q_{0,0}$ given by (also in atomic units)
\begin{equation}
q_{0,0}=\dfrac{1}{2}+\sqrt{\dfrac{1}{4}+2\mu\beta},
\end{equation}
while, for the first excited rotational state in the vibrational ground state ($n=0$, $l=1$) the corresponding wave function is also given by (\ref{Q0_normalizada}) with $q_{0,1}$
\begin{equation}
q_{0,1}=\dfrac{1}{2}+\sqrt{\dfrac{9}{4}+2\mu\beta}.
\end{equation}
\subsection{B. Vibrational excited wave functions}

To obtain vibrational excited state wave functions, we consider the effect of  $\left(T_{+}-T_{3} \right)$ on an arbitrary state $\ket{Q,q_{n,l}}$, i.e.,

\begin{small}

\begin{equation}
\hspace{-5.5cm}\nonumber \left( T_{+}-T_{3}\right)\ket{Q,q_{n,l}}=
\end{equation}
\begin{equation}
\hspace{1cm}\sqrt{-Q+q_{n,l}(q_{n,l}+\hbar)}\ket{Q,q_{n+1,l}}-q_{n,l}\ket{Q,q_{n,l}}.
\end{equation}
\end{small}

\hspace{-0.4cm} Then, using  $Q=q_{0,l}(q_{0,l}-\hbar)$ and $q_{n,l}=q_{0,l}+n\hbar$,
\begin{small}
\begin{equation}
\hspace{-4cm}\nonumber \left(T_{+}-T_{3}+q_{0,l}+n\hbar \right)\ket{Q,q_{n,l}}=
\end{equation}
\begin{equation}
\hspace{2cm}\sqrt{\hbar \left( 2q_{0,l}+n\hbar \right)(n+1) }\ket{Q,q_{n+1,l}},
\end{equation}
\end{small}
and expressing $T_{+}$ and $T_{3}$ in terms of $R$ and $P$, we obtain
\newpage
\begin{small}

\begin{equation}
\hspace{-4cm}\nonumber \left( -R+i RP+q_{0}+n \hbar\right)\ket{Q,q_{n}}=
\end{equation}
\begin{equation}
\hspace{3cm}\sqrt{\hbar \left( 2q_{0,l}+n\hbar \right)(n+1) }\ket{Q,q_{n+1}}.
\end{equation}

\end{small}

\hspace{-0.3cm}Projecting the above equation on the position basis and using the relationship between $(R,P)$ and $(r,p_r)$, we find

\begin{small}

\begin{equation}
\hspace{-3cm}\nonumber \dfrac{d f_{n,l}(r)}{dr}+\left(\dfrac{q_{0}+n \hbar}{\hbar r}-\dfrac{1}{\hbar \sigma} \right)f_{n,l}(r) =
\end{equation}
\begin{equation}
\hspace{4cm}\sqrt{\dfrac{(2q_{0,l}+n\hbar)(n+1)}{\hbar}} \dfrac{f_{n+1,l}(r)}{r} .\label{excited_state}
\end{equation}
\end{small}
\hspace{-0.2cm}Eq. (\ref{excited_state}) is a recurrence relation which relates the $nth$-vibrational wave function $Q_{n,l}(r)$ and its derivative with the following $Q_{n+1,l}(r)$. In particular, for the first vibrational excited state we take $n=0$ in (\ref{excited_state}) (taking $\hbar=1$),
\begin{equation}
\dfrac{d}{dr}\left[r Q_{0,l}(r) \right]+\left(q_{0,l}+\dfrac{ \mu \alpha }{q_{0,l}}r \right)Q_{0,l}(r)=\sqrt{2q_{0,l}} Q_{1,l}(r),
\end{equation}
and using (\ref{Q0_normalizada}), we have finally
\begin{equation}
Q_{1,l}(r)= \sqrt{\dfrac{2}{q_{0,l}}}\left( q_{0,l}+\dfrac{ \mu \alpha }{q_{0,l}}r  \right) Q_{0,l}.
\end{equation}

\section{VII. Physical insight of $SO(2,1)$ symmetry}
\subsection{A. Adjoint representation of $\mathbf{so(2,1)}$ Lie algebra}

There is a close relationship between symmetry transformations of $SO(2,1)$ Lie group and rotations in Euclidean 3D space. Also, from the \textit{adjoint representation} of $\mathbf{so(2,1)}$ Lie algebra is possible to recover the matrix expressions for the symmetry transformations of the corresponding $SO(2,1)$ Lie group. On the other hand, it is well known that these matrix representations are written in terms of hyperbolic functions, like Lorentz transformations in the case of special relativity, which allows to reveal the striking relationship between the elements of this group and the elements of $SO(3)$ Lie group.
In order to give the physical insight of $SO(2,1)$ symmetry first, we need to know how to transform the components of an observable under transformations of this Lie group, and from this, to find commutation relations between the generators of $\mathbf{so(2,1)}$ Lie algebra and the cartesian coordinates operators. Then, it is known that the components of $\textbf{r} $ changes according to
\begin{equation}
e^{-\xi \textbf{T}\cdot \hat{\textbf{n}}}r_{i}e^{\xi \textbf{T}\cdot \hat{\textbf{n}}}=\sum_{j}R_{ij}r_{j},\label{transformation}
\end{equation}
where $\xi$ is a parameter that characterizes to the elements of the Lie group, $R_{ij}$ are the elements of its matrix representation, and $\mathbf{T}$ is the vector operator defined as
\begin{equation}
\mathbf{T} \equiv T_1\hat{\mathbf{x}}+T_2\hat{\mathbf{y}}+T_3\hat{\mathbf{z}}
\end{equation}
In the case of Lorentz transformations the parameter $\xi$ corresponds to the rapidity of the \textit{boosts}, while in the $SO(3)$ group this parameters represent the angle of rotation. Here, however, $\xi$ is only an algebraic parameter. Then, for infinitesimal transformation, i.e. when $\xi\rightarrow \varepsilon<<1$, the elements of the group can be written as 
\begin{equation}
e^{\xi \textbf{T}\cdot \hat{\textbf{n}}}=1+\varepsilon \textbf{T} \cdot \hat{\textbf{n}}+\mathcal{O}(\varepsilon^{2}).\label{exponential_transformation}
\end{equation}
With this, Eq. (\ref{transformation}) becomes
\begin{equation}
r_{i}+\varepsilon\left[r_{i},\textbf{T}\cdot \hat{\textbf{n}} \right]= \sum_{j}R_{ij}r_{j}.\label{commutators_cartesian1}
\end{equation}
On the other hand, considering the adjoint representation of the $\mathbf{so(2,1)}$ Lie algebra in which the generators $T_1$, $T_2$ and $T_3$ has the matrix representation
\begin{equation}
T_{1}= \begin{bmatrix}
0 & 0 & 0 \\
0 & 0 & -1 \\
0 & -1 & 0
\end{bmatrix},
\hspace{0.3cm} T_{2}=\begin{bmatrix}
0 & 0 & 1 \\
0 & 0 & 0 \\
1 & 0 & 0
\end{bmatrix},
\end{equation}
\begin{equation}
T_{3}=\begin{bmatrix}
0&-1&0\\
1&0&0\\
0&0&0
\end{bmatrix}.
\end{equation}
and considering that these elements of the Lie algebra corresponds to the infinitesimal transformations of the corresponding Lie group, i.e.,
\begin{equation}
R_{ij}=\left( 1+\varepsilon \textbf{T} \cdot \hat{\textbf{n}}+\mathcal{O}(\varepsilon^{2})\right)_{ij} \label{Rij}
\end{equation}
we obtain for finite transformations of $SO(2,1)$:
\begin{equation}
R(\varepsilon;\hat{\mathbf{x}})=\begin{bmatrix}
1&0&0\\
0&1&-\varepsilon \\
0&-\varepsilon&1
\end{bmatrix},
\hspace{0.3cm} R(\varepsilon;\hat{\mathbf{y}})=\begin{bmatrix}
1&0&\varepsilon\\
0&1&0\\
\varepsilon&0&1
\end{bmatrix},
\end{equation}
\begin{equation}
R(\varepsilon; \hat{\mathbf{z}})=\begin{bmatrix}
1&-\varepsilon&0\\
\varepsilon&1&0\\
0&0&1
\end{bmatrix}.
\end{equation}
where we taken $\hat{\mathbf{n}}=\hat{\mathbf{x}},\hat{\mathbf{y}},\hat{\mathbf{z}}$ in Eq. (\ref{Rij}). Then, for each component of $\mathbf{r}$, we obtain, through Eq. (\ref{commutators_cartesian1}), the following commutation relations  between generators of $\mathbf{so(2,1)}$ and the cartesian coordinates:
\begin{equation}
\left[T_{1},x \right]=0,
\hspace{0.2cm}\left[T_{1},y \right]=z,
\hspace{0.2cm}\left[T_{1}, z \right]=y,
\end{equation}
\begin{equation}
\left[T_{2},x\right]=-z,
\hspace{0.2cm}\left[T_{2},y\right]=0,
\hspace{0.2cm}\left[T_{2},z\right]=-x, 
\end{equation}
\begin{equation}
\left[T_{3},x\right]=y,
\hspace{0.2cm}\left[T_{3},y\right]=-x,
\hspace{0.2cm}\left[T_{3},z\right]=0.\label{conmutator_T3}
\end{equation}
Here, by simplicity, we have performed the calculations without the imaginary constant $i\hbar$, but in the next section we recover the physical notation for commutators.

\subsection{B. A pedagogical example: selection rules for vibrational quantum number}
Considering the commutation relations between $T_{3}$ and the respective cartesian coordinates given by (\ref{conmutator_T3}), it is possible to obtain selection rules for vibrational quantum number $n$. Starting with $\left[T_{3},z\right]$ and taking the matrix element between the states $\bra{n'}$ and $\ket{n}$, with $\ket{n}\equiv\ket{Q,q_{n,l}}$,
\begin{equation}
\bra{n'}\left[T_{3},z\right]\ket{n}=0,
\end{equation}
and expanding the commutator, we obtain
\begin{equation}
(q_{n',l}-q_{n,l})\bra{n'}z\ket{n}=0.
\end{equation}
Thus, $\bra{n'}z\ket{n}=0$ unless $n'=n$.

Next, from $\left[T_{3},x\right]=i\hbar y$ and taking the same matrix element:
\begin{equation}
\bra{n'}\left[T_{3},x\right]\ket{n}=i\hbar \bra{n'}y\ket{n},
\end{equation}
and then, 
\begin{equation}
\left(q_{n',l}-q_{n,l}\right)\bra{n'}x\ket{n}=i\hbar \bra{n'}y\ket{n}.\label{selection1}
\end{equation}
In the same way, using $\left[T_{3},y\right]=-i\hbar x$,
\begin{equation}
\bra{n'}\left[T_{3},y\right]\ket{n}=-i\hbar \bra{n'}x\ket{n},
\end{equation}
and expanding,
\begin{equation}
\left(q_{n',l}-q_{n,l}\right)\bra{n'}y\ket{n}=-i \hbar \bra{n'}x\ket{n}.\label{selection2}
\end{equation}
Finally, combining Eqs. (\ref{selection1}) and (\ref{selection2}), we have
\begin{equation}
\left(q_{n',l}-q_{n,l}\right)\bra{n'}x\ket{n}=\hbar^2\bra{n'}x\ket{n}.
\end{equation}
Then, if $\bra{n'}x\ket{n}\neq 0$, we find that
\begin{equation}
q_{n',l}-q_{n,l}=\pm\hbar,
\end{equation}
which means that the eigenvalues of $T_3$ can only differ in one unit of $\hbar$.
Therefore, no transitions between different vibrational states occur unless
\begin{equation}
\Delta n =0, \pm 1. \label{selection rules}
\end{equation}

Eq. (\ref{selection rules}) corresponds to the selection rules for the vibrational states of the molecules, which is a known result for molecular spectroscopy and quantum chemistry, but here we have obtained it from a entirely algebraic approach.

\section{VIII. Conclusions and remarks}
In this work the usefulness of the $\mathbf{so(2,1)}$ Lie algebra in the solution of rovibrational Schr\"odinger equation for Kratzer oscillator has become evident. Indeed, when comparing the solution method provided by the Lie algebra approach with the standard resolution of Schr\"odinger equation for this oscillator model through hypergeometric series \cite{matamala}, it is possible to note the simplicity with which we have obtained the energies and the states of this system. Moreover, identifying the vibrational ground state with the lower bound of this Lie algebra, and representing its generators in the position basis, it has been possible to decrease the degree of the Schr\"odinger equation obtaining a first order differential equation whose resolution is considerably simpler than the preceding equation. Although there are other works that employ Lie algebras in order to solve the Schr\"odinger equation for molecular oscillators \cite{matamala1}, there is no presence concerning a clear interpretation about the physical meaning of the generators of the algebra.
	
Finally, a good challenge for a prospective work would be generalizing the $\mathbf{so(2,1)}$ Lie algebra by means of a larger algebraic structure that contains ladder operators for the angular quantum number, and also, diagonal ladder operators which changes both $l$ and $n$ quantum numbers simultaneously.
\section{Acknowledgments}
PhD Adelio R. Matamala is thanked for his inspiring ideas about the applications of Lie algebraic theory to the molecular physics and theoretical chemistry. Mr. Bruno Limbardo are thanked for his insightful comments concerning the preparation of this manuscript. 
PhD (c) Carmen Nabalón V. for her linguistic advice concerning the writing of this manuscript.



\end{document}